\documentclass{article}

\usepackage{arxiv}

\usepackage[utf8]{inputenc} 
\usepackage[T1]{fontenc}    
\usepackage{hyperref}       
\usepackage{url}            
\usepackage{booktabs}       
\usepackage{amsfonts}       
\usepackage{nicefrac}       
\usepackage{microtype}      
\usepackage{lipsum}		
\usepackage{graphicx}
\usepackage[numbers]{natbib}
\usepackage{doi}

\usepackage{multirow}
\usepackage{longtable}
\usepackage[font=footnotesize]{caption}
\usepackage{subcaption}
\usepackage{xcolor}
\usepackage{algpseudocode}
\NewDocumentCommand{\codeword}{v}{
\texttt{\textcolor{blue}{#1}}}

\usepackage{listings}

\usepackage{xparse}

\title{\textbf{CMDA}: a tool for Continuous Monitoring Data Analysis}

\author{ \href{https://orcid.org/0000-0000-0000-0000}{\includegraphics[scale=0.06]{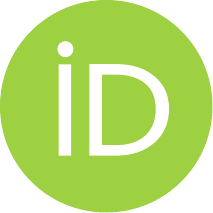}\hspace{1mm}Pejman Farhadi Ghalati} \\
	Joint Research Center for Computational Biomedicine\\
	RWTH Aachen University\\
	Aachen, Germany \\
	\texttt{farhadi@combine.rwth-aachen.de} \\
	\And
	{Andreas Schuppert} \\
	Joint Research Center for Computational Biomedicine\\
	RWTH Aachen University\\
	Aachen, Germany \\
	\texttt{schuppert@combine.rwth-aachen.de} \\
}

\renewcommand{\shorttitle}{\textbf{CMDA} Package}

\hypersetup{
pdftitle={CMDA Package},
pdfsubject={cs.CE},
pdfauthor={Pejman Farhadi Ghalati, Andreas Schuppert},
pdfkeywords={time series, feature extraction, preprocessing},
}

\begin{document}
\maketitle

\begin{abstract}
Over the last few years, with the growth of time-series collecting and storing, there has been a great demand for tools and software for temporal data engineering and modeling. This paper presents a generic workflow for time series data research, including temporal data importing, preprocessing, and feature extraction. This framework is developed and built as a robust and easy-to-use Python package, called CMDA, with a modular structure that offers tools to prepare raw data, allowing both scientists and non-experts to analyze various temporal data structures.
\end{abstract}


\section{Introduction}
Data generation and collection in real-time is more ubiquitous than ever due to technological progress in sensors, transmission, and data storage. This has resulted in enormous datasets, namely Big Data, in various fields such as engineering, medical, physical, and social sciences. While effective modeling and simulation of such datasets can offer rich information for interpretation, prediction, monitoring, training, and decision making, their scope and complexity create significant design and implementation issues \cite{frohlich_hype_2018}.\\
For instance, in the healthcare systems, continuous monitoring data has developed into a breakthrough for health diagnosis, subsequent treatment, and proactive health tracking \cite{athavale_biosignal_2017}. The recent advancements in wearable technology and monitoring systems have offered unseen opportunities for learning mechanisms of biomedical processes from temporal data as they provide data on physiological occurrences that indicate an individual’s health and comfort. 
However, it must be taken into account that, the analysis of continuous monitoring data is highly affected by data quality issues and even complex monitoring devices can monitor only a small subset of medical parameters involved in the process under consideration. Furthermore, the required analysis strategies for monitoring data vary significantly between monitoring a few parameters assessed on dense sampling rates (e.g., ECG, EEG, or glucose data) and monitoring several parameters over hours with low sampling frequency, as in ICU. 
On account of this, there is a significant demand for robust and easy-to-implement tools to facilitate continuous data wrangling and feature engineering.
In order to address this, we introduce the \textbf{CMDA} package to provide a comprehensive monitoring data analysis workflow.
CMDA, developed and built as a Python package, is a standard framework for time series data research, including temporal data importing, preprocessing, and feature extraction.

\section{Structure}
\begin{figure}[ht]
\includegraphics[width=0.9\textwidth]{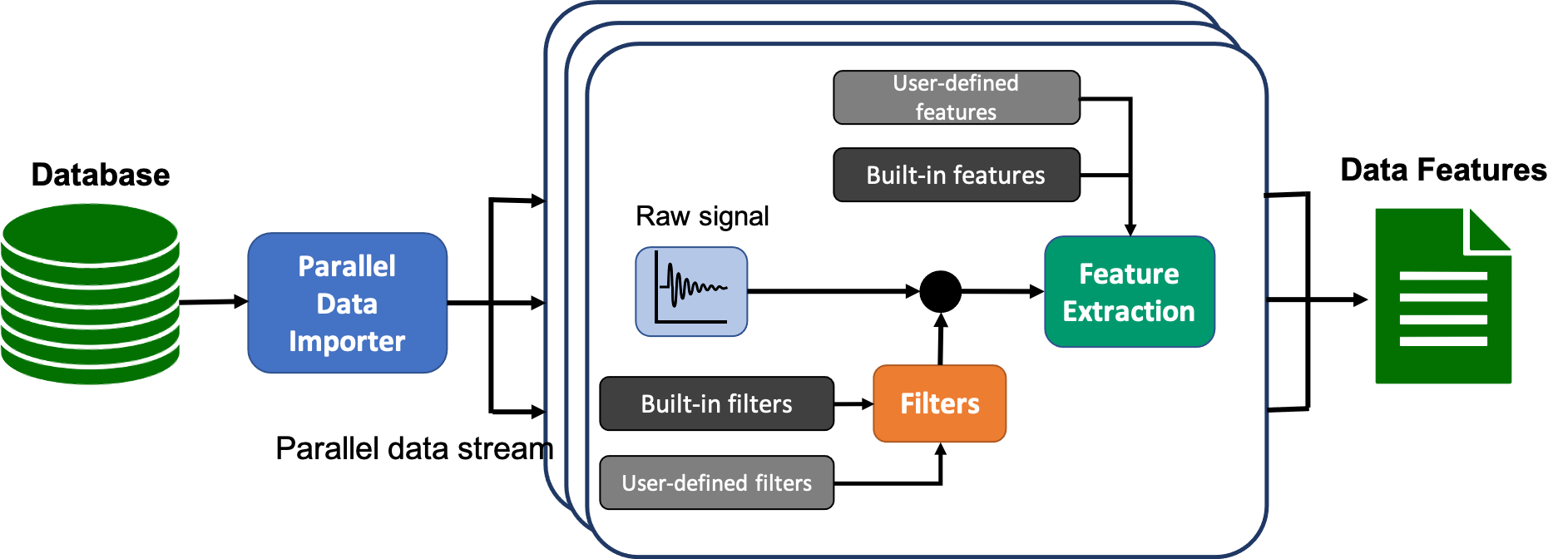}
\centering
\caption{Illustration of the CMDA workflow. The pipeline consists of three components. In the feature engineering stage, data records can be imported into the pipeline either consecutively or in parallel. Once filters are applied (if needed) and features are extracted, these data features are saved to a file. The features can be used later for any ML task.}
\label{fig:workflow}
\end{figure}

A general workflow for continuous data analysis consists of four steps: importing the time-series records, preprocessing (missing value imputation and denoising), feature extraction, and model creation.\\
Initially, continuous data is imported from various sources. Second, the imported data is preprocessed based on their quality to verify that only the relevant data reach the subsequent steps. This can involve eliminating irrelevant artifacts from the data (via filtering and outlier removal techniques) and windowing signals to a set of time bins. In the step before modeling, features relevant to a particular problem are derived from filtered data.\\
Feature extraction or engineering uncovers the masked characteristics of raw temporal data. In other words, it can be described as a set of features corresponding to the signal's specific form or pattern. Additionally, feature extraction is generally a procedure that reduces the dimensionality of an input signal or compresses its data and thus reduces the number of resources needed for analysis. Performing extraction of descriptors on a raw signal would result in low dimensional data that could accurately elaborate the input raw data and aid in developing a practical and interpretable input for machine learning tasks.\\ 
CMDA offers a set of tools for the mentioned steps in a modular structure, allowing users to use modules individually as part of another project, depending on the application. Nevertheless, the whole process can be integrated into a pipeline, which can be executed in a parallel fashion.
Figure \ref{fig:workflow} depicts the components of the feature engineering pipeline. The other noteworthy attribute of CMDA is that besides the built-in feature functions, user-defined functions can be added easily to the pipeline, which enables more adaptable use of the feature engineering module.
A detailed description of the package architecture, including the data importing modules and feature extraction modules, as well as the list of the built-in feature functions, can be found in the next section along with tutorials in appendix \ref{sec:appx}.

\section{Architecture}
\subsection{Data Importer Module}
Data importer (\codeword{cmda.import_data}), the first component in the pipeline, offers tools for loading multiple CSV files from a server or local disk into the python coding environment (\codeword{cmda.import_data.ImportCSV}). It should be taken into account that the CSV files must contain signals as columns and temporal instances as rows.\\
The importer module also supports importing DAT files directly from Physionet databases (\codeword{cmda.import_data.ImportWFDB}) using WFDB package APIs \cite{xie_chen_waveform_nodate}.
Physionet, a funded project by the National Institutes of Health, is a vast and growing archive of digital recordings of physiologic signals and corresponding data for the use of medical research \cite{goldberger_physiobank_2000}.\\
The importer functions can save computation time by parallelizing the data loading process. 
Moreover, in the case of a very long time series, if windowing or segmentation is required, a rolling window importer function performs the time binning with a defined window length and step size. It distributes the importing process for each window across available nodes in parallel. Users can access the rolling window via \codeword{cmda.import_data.RollingWindowCSV}. The library offers modules for importing continuous data from both local disk and Physionet databases.

\subsection{Preprocessing Module}
\codeword{cmda.preprocessing.Filters} is a tool for applying different filters and outlier removal functions on a numeric signal.
Users can select from several built-in filters using a dynamic attribute of the \codeword{Filters} object, \codeword{.add}, which provides access to the following functions:
\begin{longtable}
{ |c||c|p{8 cm}|}
 \hline
 Feature function& Feature & Description\\
 \hline
 \codeword{rm_outlier}   & Remove Outlier    & Removing outliers based on a $low$ and $high$ threshold\\
 \hline
 \codeword{rm_outliers_quantile}   & Remove Quantiles    & Removing the values in an array based on $low$ and $high$ defined quantiles\\
 \hline
 \codeword{butter_filter}   & Butterworth Filter    & A predesigned Butterworth filter that can be used as a band-pass denoisng tool\\
 \hline
 \codeword{interpolate}   & Interpolation    & A $linear$ or $cubic$ interpolation tool that can be used for imputing missing values in an array\\
 \hline
\caption{Built-in preprocessing features}
\label{tab:tab_filter}
\end{longtable}

Moreover, users can easily add their own defined filters to the \codeword{Filters} object instance using \codeword{.udf.add}. This functionality makes it possible to tailor the filters based on the application.

\subsection{Feature Extraction Module}
\codeword{cmda.feature_extraction.Features} is a tool for extracting multiple features from a numeric time series array.
This library provides several ready-to-use built-in time domain, spectral domain, nonlinear, and wavelet features that can be selected and added to a \codeword{Features} object instance via \codeword{.add} function.\\
Additionally, users can define and add their own functions to the \codeword{Features} object instance using \codeword{.udf.add}.
The available built-in functions in the current version of the package are described in the following sections.

\subsubsection{Time-domain Features}
Time-domain features describe signals over a span of time. The list of all built-in time-domain features is shown in Table \ref{tab:tdf}. These functions are defined using the \codeword{NumPy} \cite{harris_array_2020} package.

\begin{longtable}
{ |c||c|p{8 cm}|}
 \hline
 Feature function& Feature & Description\\
 \hline
 \codeword{mean}   & Mean    & Arithmetic mean of an array elements
\\
 \hline
 \codeword{max}   & Maximum    & Maximum value of an array elements\\
 \hline
 \codeword{min}   & Minimum    & Minimum value of an array elements\\
 \hline
 \codeword{median}   & Median    & Median value of an array elements\\
 \hline
 \codeword{std}   & Standard Deviation    & Standard deviation of an array elements\\
 \hline
 \codeword{skewness}   & Skewness    & Skewness of an array elements\\
 \hline
 \codeword{kurtosis}   & Kurtosis    & Kurtosis of an array elements\\
 \hline
 \codeword{p2p}   & Peak-to-Peak    & Peak to peak difference (maximum - minimum values) of an array elements\\
 \hline
 \codeword{rms}   & Root Means Square    & Root means square of an array elements\\
 \hline
 \codeword{zcr}   & Zero Crossing Rate    & The rate at which an 1-D array changes 
    from positive to zero to negative or from negative to zero to positive\\
 \hline
 \codeword{mad}   & Mean Absolute Deviation    &  The median of the absolute deviations an array. Also called median of slopes\\
 \hline
  \codeword{mns}   & Mean Slope    &  The mean differences of an array\\
 \hline
\caption{Built-in time-domain features}
\label{tab:tdf}
\end{longtable}

\subsubsection{Frequency-domain Features}
Frequency-domain features, extracted from the Fourier transform of a signal, describe its characteristics in the frequency domain. The list of all built-in frequency-domain features is shown in Table \ref{tab:fdf}. These functions are defined using the \codeword{NumPy} \cite{harris_array_2020} and \codeword{SciPy} \cite{2020SciPy-NMeth} packages.

\begin{longtable}
{ |c||p{4 cm}|p{6 cm}|}
 \hline
 Feature function& Feature & Description\\
 \hline
 \codeword{periodogram}   & Periodogram    & The power spectrum density of an array \cite{naik_usefulness_2012}\\
 \hline
 \codeword{welch}   & Welch Periodogram    & The power spectrum density of an array computed using welch method.\\
 \hline
 \codeword{mnf}   & Mean frequency    & The center of the distribution of power across frequencies. It is calculated as the sum of product of the power spectrum and the frequency divided by the total sum of the power spectrum \cite{naik_usefulness_2012}.\\
 \hline
 \codeword{mdf}   & Median frequency    & The frequency at which the power spectrum is divided into two parts with equal powers \cite{naik_usefulness_2012}.\\
 \hline
 \codeword{vcf}   & Variance of central frequency    & The variance of power spectrum amplitudes from its mean frequency \cite{naik_usefulness_2012}.\\
 \hline
 \codeword{peaks}   & Peak Frequency    & The frequency at which the maximum of power spectrum occurs \cite{naik_usefulness_2012}.\\
\\
 \hline
 \codeword{psr}   & Power spectrum ratio    & The ratio between the energy around the maximum value of the power spectrum and the whole energy of the power spectrum \cite{naik_usefulness_2012}.\\
 \hline
 \codeword{band_power}   & Band Power    & The power spectral density in a specified frequency band.\\
 \hline
 \codeword{band_std}   & Band Standard Deviation    & The power spectral standard deviation in a specified frequency band.
\\
 \hline
 \codeword{band_mnf}   & Band Mean Frequency    & The mean frequency of a specified frequency band.
\\
 \hline
 \codeword{band_mdf}   & Band Median Frequency    & The median frequency of a specified frequency band.
\\
 \hline
\caption{Built-in frequency-domain features}
\label{tab:fdf} 
\end{longtable}

\subsubsection{Entropy Features}
Shannon entropy and its derivations are methods for estimating the complexity of dynamic systems. Table \ref{tab:entropy} shows the available entropy built-in functions in \textbf{CMDA}.

\begin{longtable}
{ |c||c|p{8 cm}|}
 \hline
 Feature function& Feature & Description\\
 \hline
 \codeword{entropy}   & Shannon Entropy    & A technique for estimating the irregularity in temporal data \cite{richman_physiological_2000}\\
 \hline
 \codeword{sample_entropy}   & Sample Entropy    & A technique for estimating the complexity of dynamic physiological systems based on the existence of patterns \cite{richman_physiological_2000}\\
 \hline
 \codeword{perm_entropy}   & Permutation Entropy    & A derivation of Shannon entropy, used to assess the complexity of time series data, proposed by Bandt\cite{bandt_permutation_2002}. It quantifies the complexity of a system based on similar ordinal patterns.\\
  \hline
 \codeword{spectral_entropy}   & Spectral Entropy    & A measure of the uniformity of a power spectrum frequency distribution \cite{pan_spectral_2009}.\\
 \hline
\caption{Built-in entropy features}
\label{tab:entropy}
\end{longtable} 

\subsubsection{Wavelet Features}
The wavelet transform provides a significant method for joint time-frequency feature extraction. It was developed to deal with non-stationary signals. The wavelet transform is classified as discrete or continuous. Computing wavelet coefficients at every feasible scale can be time-consuming and produce a large amount of data. As a result, the DWT (Discrete wavelet transform) is extensively used. However, it suffers from the downsample resolution of DWT at higher decomposition levels and destroys the time invariance property. The Stationary wavelet transform (SWT) \cite{combes_real-time_1990} is a wavelet transform algorithm designed to overcome the downsample problem of DWT and preserves the time information. CMDA offers a built-in \codeword{swt_features} function that returns both time-domain and frequency-domain coefficients from decomposition levels. The following parameters can be given to the function:
\begin{itemize}
    \item Mean Frequency (mnf)
    \item Power Spectrum Ratio (psr)
    \item Peak Frequency (peak)
    \item Median Absolute Deviation (mds)
    \item Mean Slope (mns)
    \item Shannon Entropy (see)
    \item Permutation Entropy (perm-ent)
    \item Normalized Energy (nse)
\end{itemize}
The explanation of each parameter can be found in Tables \ref{tab:tdf}, \ref{tab:fdf}, and \ref{tab:entropy}.

\subsection{Feature Engineering Pipeline}
\codeword{cmda.pipeline.Pipeline} is a tool for importing, pre-processing and feature extraction of multiple records (files) in an automated manner.\\
Users must create an importer object instance using \codeword{cmda.import_data}, a feature object instance using \codeword{cmda.preprocessing.Features}, and, if needed, a filter object instance with \codeword{cmda.preprocessing.Filters}. They must then feed these instances into the pipeline. The process can be executed either consecutively or in parallel using the \codeword{.run} function.

\section{Discussions}
The CMDA Python package is designed to facilitate a monitoring data analysis workflow, with functions ranging from reading multiple datasets to denoising and feature extraction. Its modular design allows for implementation across various applications and supports the integration of user-defined components. The package provides a fast and straightforward workflow with minimum coding requirements, making it feasible for the researchers to prepare ready-to-analyze extensive monitoring datasets before applying any sophisticated strategy. 
In the future development of CMDA, we aim to add more built-in features and pre-processing functions to cover broader applications. In addition, embedding unsupervised feature engineering methods such as deep autoencoders and signal to image transformation is another objective of the package expansion. Until then, we hope CMDA enhances healthcare data expertise by setting a standard for monitoring data analysis.

\section{Software Availability}
The CMDA toolkit is freely accessible as open-source on GitHub. The software can be accessed and downloaded from:

\href{https://github.com/JRC-COMBINE/cmda}{www.github.com/JRC-COMBINE/cmda}

A README file within the repository provides insights into the software's setup process, required dependencies, and instructions for use. For any concerns or queries, please use the "Issues" section of the GitHub repository. Community input and contributions to improve the toolkit are highly welcomed.

\bibliographystyle{abbrv}
\bibliography{references}  





\appendix
\section{Examples}\label{sec:appx}

\definecolor{codegreen}{rgb}{0,0.6,0}
\definecolor{codegray}{rgb}{0.5,0.5,0.5}
\definecolor{codepurple}{rgb}{0.58,0,0.82}
\definecolor{backcolour}{rgb}{0.95,0.95,0.92}

\lstdefinestyle{mystyle}{
  backgroundcolor=\color{backcolour}, commentstyle=\color{codegreen},
  keywordstyle=\color{magenta},
  numberstyle=\tiny\color{codegray},
  stringstyle=\color{codepurple},
  basicstyle=\ttfamily\footnotesize,
  breakatwhitespace=false,         
  breaklines=true,                 
  captionpos=b,                    
  keepspaces=true,                 
  numbers=left,                    
  numbersep=5pt,                  
  showspaces=false,                
  showstringspaces=false,
  showtabs=false,                  
  tabsize=2
}

\lstset{style=mystyle}

\subsection{Feature Object} \label{ssec:feature_obj}
\codeword{cmda.feature_extraction.Features} is a tool for extracting multiple features at the same time from a numeric signal \codeword{x}. One can select from a wide variety of built-in features, as well as adding user-defined features. This functionality makes it possible to tailor the features based on the application.
The following subsections show examples of time domain, frequency domain, and user-defined feature extraction.

\subsubsection{Import the data}
We start by using sample data contains an ECG signal and ABP (Arterial Blood Pressure) signal with a length of 10 seconds and 125 Hz sampling rate. The toy data can be accessed from \codeword{cmda.data.ecg_abp_sample}.

\begin{lstlisting}[language=Python, caption=Importing a sample data]
import numpy as np
import matplotlib.pyplot as plt

from cmda.data import ecg_apb_sample

data = ecg_apb_sample()

ecg = data["ECG"]
abp = data["ABP"]

fs = 125
t = np.linspace(1, 1 / fs, len(ecg))
fig, (ax1, ax2) = plt.subplots(2, sharex=True)
ax1.plot(t, ecg)
ax1.set_ylabel("ECG [mV]")
ax2.plot(t, abp, c="red")
ax2.set_ylabel("ABP [mmHg]")
ax2.set_xlabel("Time [seconds]")
plt.show()
\end{lstlisting}

Output:
\begin{figure}[h]
\includegraphics[width=0.8\textwidth]{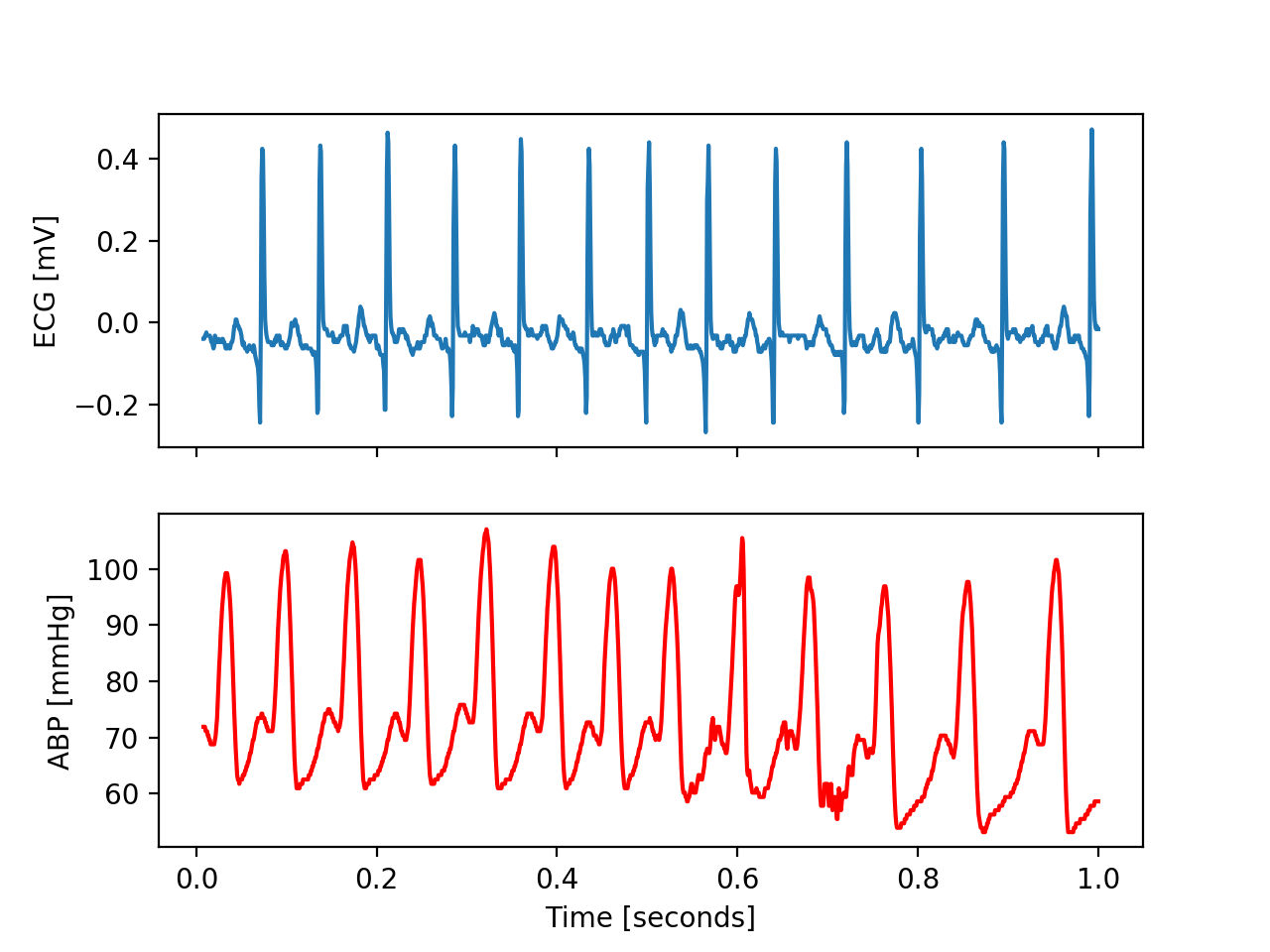}
\centering
\caption{Imported ECG and ABP sample signals}
\label{fig:pipeline}
\end{figure}

\subsubsection{Time-domain feature extraction}
In this example, we extract some built-in time-domain features of the ABP signal. To start, we must call the \codeword{cmda.feature_extraction}.Features object and add the desired features using the \codeword{add} function.\\
After adding the features, we apply the \codeword{transform} function to the signal and get the features. The \codeword{transform} function returns the extracted features as a python dictionary.

\begin{lstlisting}[language=Python, caption=Time-domain feature extraction]
from cmda.feature_extraction import Features
from cmda.data import ecg_apb_sample

data = ecg_apb_sample()
x = data["ABP"]

feature = Features()

# add built-in time-domain features
feature.add.mean()
feature.add.max()
feature.add.min()
feature.add.median()
feature.add.skewness()
feature.add.kurtosis()
feature.add.std()
feature.add.p2p()
feature.add.zcr(center=True)

feature.transform(x=x, fs=125)
\end{lstlisting}

Output:
\begin{lstlisting}[caption=Time-domain feature extraction output]
{'mean': 72.385,
 'max': 107.03125,
 'min': 53.125,
 'median': 69.53125,
 'skewness': 0.9154627781784881,
 'kurtosis': -0.15299373030467667,
 'std': 13.268216310510619,
 'p2p': 53.90625,
 'zcr': 42}
\end{lstlisting}

\subsubsection{Frequency-domain feature extraction}
In this part, we extract some built-in frequency-domain features of the ECG signal. Similar to the previous example, we call the \codeword{cmda.feature_extraction.Features} object, and append the desired features using the \codeword{add} function.\\
Using the \codeword{transform} function returns the extracted features as a python dictionary.
\begin{lstlisting}[language=Python, caption=Frequency-domain feature extraction]
from cmda.feature_extraction import Features
from cmda.data import ecg_apb_sample

data = ecg_apb_sample()
x = data["ECG"]

feature = Features()

# add frequency-domain built-in features
feature.add.mnf(spectrum="ps")
feature.add.mdf(spectrum="ps")
feature.add.stdf(spectrum="ps")
feature.add.psr(spectrum="welch", int_limit_ratio=0.01)
feature.add.peaks(spectrum="welch", n_peaks=1, height=True, width=True)
feature.add.band_power(spectrum="ps", low=1, high=7)
feature.add.band_mnf(spectrum="ps", low=1, high=7)

feature.transform(x=x, fs=125)
\end{lstlisting}

Output:
\begin{lstlisting}[caption=Frequency-domain feature extraction output]
{'mnf': 9.857968968908676,
 'mdf': 9.0,
 'stdf': 6.657686374206612,
 'psr_0.01': 0.1086823980353686,
 'peak_freq_1': 3.90625,
 'peak_height_1': 0.0009058970155242925,
 'peak_width_1': 4.386465062242469,
 'power_[1,7]Hz': 0.3921784430303598,
 'mnf_[1,7]Hz': 4.277888519896743}
\end{lstlisting}

\subsubsection{User-defined feature extraction}
In this section, we derive some features, including a user-defined feature from the ABP signal. Similar to the previous example, we call the \codeword{cmda.feature_extraction.Features} object, Define the feature and add them to the feature object, using \codeword{add_udf} function. The \codeword{transform} function returns the extracted features as a python dictionary.

\begin{lstlisting}[language=Python, caption=User-defined feature extraction]
import numpy as np
from cmda.feature_extraction import Features
from cmda.data import ecg_apb_sample

data = ecg_apb_sample()
x = data["ABP"]

# Define the feature
# Please take note, that the output must be a dictionary
# ,where the keys are the extracted features
def quantile(x,low = 0.1,high =0.9):
    q_low = np.quantile(x,q=low)
    q_high = np.quantile(x,q=high)
    
    res = {"q_low":q_low, "q_high":q_high}
    return res


# Call the Feature object
feature = Features()

# add built-in features
feature.add.mean()
feature.add.max()
feature.add.zcr(center=True)
feature.add.mnf(spectrum="ps")

# add user-defined feature
feature.udf.add(quantile)

feature.transform(x=x, fs=125)
\end{lstlisting}

Output:
\begin{lstlisting}[caption=User-defined feature extraction output]
{'mean': 72.385,
 'max': 107.03125,
 'zcr': 42,
 'mnf': 2.0056974184541025,
 'q_low': -13.79125,
 'q_high': 23.70875}
\end{lstlisting}

\subsection{Multivariate Feature Extraction}

In this tutorial, we show how to extract features from a multivariate signal. In the first example, we extract similar features from all the variables, and in the second part, we show how to extract different features from each variable. Please read \ref{ssec:feature_obj} on how to create a feature object first. We use the same data as used in the Feature Object tutorial (\ref{ssec:feature_obj}). The data contain two signals (ECG, ABP) with a sampling frequency = 500.

\subsubsection{Feature extraction with one feature object}
To extract features from a dataset, we use \codeword{cmda.feature_extraction.extract_features} function. First, we create a feature object with \codeword{cmda.feature_ectraction.Features}.

\begin{lstlisting}[language=Python, caption=Multivariate feature extraction using one feature object]
from cmda.feature_extraction import Features
from cmda.feature_extraction import extract_features
from cmda.data import ecg_apb_sample

# load the data
data = ecg_apb_sample()

# Create the feeature object
features = Features()

# Add built-in features
features.add.std()
features.add.mnf()
features.add.stdf()

# extract eatures
res = extract_features(data=data, feature_obj=features)

print(res)
\end{lstlisting}

Output:
\begin{lstlisting}
{'ECG_std': 0.0912508348533301,
 'ECG_mnf': 0.0788637517512694,
 'ECG_stdf': 0.053261490993652905,
 'ABP_std': 13.268216310510619,
 'ABP_mnf': 0.01604557934763282,
 'ABP_stdf': 0.012480758701313835}
\end{lstlisting}

\subsubsection{Feature extraction with multiple feature objects}
In various real applications, extracting similar features from different signals may lead to wrong data compression and unnecessary information. To avoid that, we show how to create and apply signal-specific feature objects.

\begin{lstlisting}[language=Python, caption=Multivariate feature extraction using one feature object]
from cmda.feature_extraction import Features
from cmda.feature_extraction import extract_features
from cmda.data import ecg_apb_sample

# load the data
data = ecg_apb_sample()

# Create the feature object for ECG signal
ecg_features = Features()

# Add built-in features
ecg_features.add.mnf()
ecg_features.add.stdf()


# Create the feature object for ECG signal
abp_features = Features()

# Add built-in features
abp_features.add.mean()
abp_features.add.std()
abp_features.add.max()
abp_features.add.min()

# Create the feature object as a dict
features = {'ECG':ecg_features, 'ABP':abp_features}

# extract eatures
res = extract_features(data=data, feature_obj=features)

print(res)
\end{lstlisting}

Output:
\begin{lstlisting}
{'ECG_mnf': 0.0788637517512694,
 'ECG_stdf': 0.053261490993652905,
 'ABP_mean': 72.385,
 'ABP_std': 13.268216310510619,
 'ABP_max': 107.03125,
 'ABP_min': 53.125}
\end{lstlisting}

\subsection{Feature Extraction Pipeline}
This tutorial shows how to implement \codeword{cmda.pipeline.Pipeline} to extract features from a Physionet WFDB.
We extract features from the ECGRDVQ database. The database contains 5232 extracted 10-second standard 12 lead ECG segments. The meta-data of the database (\codeword{ecgrdvq_metadata}), which is a dictionary consisting of the record names \codeword{record_name} and their corresponding public directories \codeword{public_dir}, is available in \codeword{cmda.data.wfdb}. The pipeline includes importing the data, applying filters to the records, and feature extraction from records. To import the data from Physionet WFDB, we use the \codeword{cmda.read_data.ReadWFDB}. We create the filtering object by \codeword{cmda.filter.Filters} and create the feature object by \codeword{cmda.feature_extraction.Features}. A detailed tutorial on creating feature objects can be found in \ref{ssec:feature_obj}.

\begin{lstlisting}[language=Python, caption=Multivariate feature extraction using one feature object]
from cmda.data.wfdb import ecgrdvq_metadata
from cmda.read_data import ReadWFDB
from cmda.preprocessing import Filters
from cmda.feature_extraction import Features
from cmda.pipeline import Pipeline

# Load the meta-data
metadata = ecgrdvq_metadata()

# get the record names and their corresponding public directories
record_names = metadata['record_name']
public_dir = metadata['public_dir']

# Build the importer object using ReadWFDB 
# Set the channels to ["II",'V1'] to import these channels exclusively. 
importer = ReadWFDB(record_names=record_names, public_dir=public_dir, channels=['II','V1'])

# Create the filter object and add a buuterworth low-pass filter
filter_obj = Filters()
filter_obj.add.butter_filter(cutoff=60, btype="lowpass")

# Create the feature object
feature_obj = Features()
feature_obj.add.mnf()
feature_obj.add.mdf()
feature_obj.add.psr()
feature_obj.add.stdf()
feature_obj.add.peaks(n_peaks=1,spectrum='welch',height=False,width=False,nperseg=512)
feature_obj.add.band_power(low=0.6, high=2,spectrum='welch')
feature_obj.add.band_power(low=2, high=4,spectrum='welch')
feature_obj.add.band_power(low=4, high=6,spectrum='welch')
feature_obj.add.band_power(low=6, high=10,spectrum='welch')
feature_obj.add.band_power(low=10, high=15,spectrum='welch')
feature_obj.add.band_power(low=15, high=30,spectrum='welch')

# Build the pipeline
ecgrdvq_pipeline = Pipeline(importer=importer,features=feature_obj, filters=filter_obj)

# Run the pipeline
# Set the dataframe_output to True, to get the extracted as a Pandas dataframe
# n_jobs set the number of cpu cores for parallel feature extraction
res = ecgrdvq_pipeline.run(n_jobs=4, dataframe_output = True)
\end{lstlisting}

Output:
\begin{lstlisting}
Running the pipeline on 5232 instances...

100% [|||||||||||||||] 5232/5232 [00:05<00:00,  2.23it/s]

finished!
\end{lstlisting}

The required time for parallel running the above pipeline was around 2640 seconds for a PC with a 3,5 GHz Dual-Core Intel Core i7 processor and 16 GB 2133 MHz LPDDR3 memory. The process included retrieving the data from Physionet data servers too.

\end{document}